# Tunneling Spectroscopy of Quantum Hall States in Bilayer Graphene PN Networks

Ke Wang[1], Achim Harzheim[1], Ji Ung Lee[2], Takashi Taniguchi[3], Kenji Watanabe[3], Philip Kim[1*]

[1]Department of Physics, Harvard University, Cambridge, 02138, MA, USA

[2]Colleges of Nanoscale Science and Engineering, SUNY Polytechnic Institute, Albany, 12203, NY, USA

[3]National Institute for Materials Science, Namiki 1-1, Ibaraki 305-0044, Japan.

*Correspondence to: pkim@physics.harvard.edu

**Abstract**: Two dimensional electronic systems under strong magnetic field form quantum Hall (QH) edge states, which propagate along the boundary of a sample with a dissipationless current. Engineering the pathway of these propagating one-dimensional chiral modes enables the investigation of quantum tunneling between adjacent QH states. Here, we report tunneling transport in spatially controlled networks of QH edge states in bilayer graphene. We observe resonant tunneling between co-propagating QH edges across barriers formed by electrically defining incompressible strips. Employing spectroscopic tunneling measurements enable the direct probing of the spatial profile, density of states, and compressibility of the QH edge states with an unprecedented energy resolution. The capability to engineer the QH edge network provides an opportunity to build future quantum electronic devices supported by rich underline physics.

**One Sentence Summary:** We present spectroscopic tunneling measurements of gate-defined quantum Hall states in a bilayer graphene PN junction network, allowing us to probe fundamental physical properties of QH states.

**Main Text:**

Recent advances in the quality GaAs/AlGaAs *(1,2)* and graphene *(3,4,5)* have significantly improved the electronic properties of two-dimensional electron gases (2DEGs), where integer states and non-Abelian fractional states appear from exotic many-body excitations *(6, 7)*. These states hold a fundamental interest both in condensed matter physics as well as possible applications towards topological quantum computing *(7)*. By controlling the spatial location of these unconventional quantum Hall (QH) edge states and studying their characteristic energies, one can enable new quantum electronic devices.

The energy gaps and electronic compressibility of QH states have been addressed by several experimental techniques, such as single electron transistors *(8,9)*, optical spectroscopy *(10)* and capacitive measurement techniques *(11)*. However, while these techniques provide a direct access to the bulk properties, the spectroscopic probe of carrier transport mediated QH edge states has yet to be realized.

The structure of chiral one-dimensional (1D) QH edges has been studied using scanning probe microscopy *(12,13)*. In these studies, the existence of electrically conducting compressible strips (CS) separated by electrically insulating incompressible strips (IS), running along the rising potential of the sample's edge, has been identified. For filling fraction ν in the quantum Hall regime, Landau gaps are directly related to the width of the corresponding CS and IS *(14)*. With electron-electron interaction, it has been shown that CS and IS can be reconstructed into fractionalized 1D modes, including the neutral charge mode *(15)*. In order to characterize the spatial and energetic characteristics of the QH edge states, tunneling measurements are often employed *(16)*. In conventional quantum Hall devices, however, adjacent edge states tend to share the same chemical potential as the current-injecting contact, and consequently, the effective tunneling bias across any incompressible strip vanishes. Chemical potential bias across two sets of QH states can be established using gate-defined quantum point contacts (QPC) *(17)*. However, the tunneling widths near QPCs are often not uniformly defined, and the QH states are without a well-defined IS region between them, obscuring the spectroscopic resolution during tunneling measurements.

In this work, we report transport measurements of a gate-defined bilayer graphene (BLG) PN network in the QH regime with a tunable bandgap *(10)*. The tunable gap provides an additional control of the QH states by allowing a tunneling bias to form across co-propagating QH states near a gate-defined PN junction. By carefully designing the PN junction network, together with optimized gate configurations, we can uniformly space-out the QH edge states to allow the chemical potential of the edge states to be addressed individually. Fig. 1A shows an optical microscope image of a BLG device with three local back gates and a top gate, which are used to form a network of PN junctions in the BLG channel. The BLG channel is encapsulated in hBN to achieve high carrier mobility (~ 30 m$^2$/Vs) and is contacted by edge contacts (Fig. 1B) (18, 19). The three horizontal bottom gates and one vertical top gate electrostatically divide the sample into nine regions. Voltages applied to these gates determine the carrier density and the bandgap Δ induced by the displacement field in each region. To maintain a finite tunnel barrier, we keep the sign of the displacement field unchanged across a PN junction. With a magnetic field, electron and hole Landau levels appear in the N and P regions separated by an insulating region (often referred as ν=0 Landau gap shown in Fig. 1C), whose gap-size remains finite even at zero magnetic fields thanks to the displacement-field-induced bandgap.

By controlling the voltages applied to the local and top gates, we can create nine segmented regions in the BLG channel (Fig. 1B-C) whose boundaries guide the corresponding QH edge states. Fig. 1B-C shows two different configurations of QH edge states. For each configuration, we keep the bottom two outer gates at $V_E$=-10 V and the top gate at $V_T$=+8 V. With this gate voltage setting, we create intrinsically gaped regions in the outer middle sections (regions #2 and #8), while keeping the 'lead' regions (#1,3,4,6,7, and 9) P-doped. We control the transport properties of the QH edge states by changing the voltage of the bottom middle gate, $V_B$, to alter the doping of region #5. For this gate, we define a reference gate voltage $\Delta V_B$ where $\Delta V_B$ = 0 corresponds to the charge neutrality point of the center region (#5). For $\Delta V_B < 0$, the middle bridge (#5) becomes P-doped. In this unipolar gate configuration, the outermost QH edge states propagate the bridge region, connecting the two sides of the device (Fig. 1B). The number of connecting QH edge states corresponds to the filling fraction of the middle bridge (#5), ν. As $V_B$ increases and the chemical potential of the bridge region passes through the charge neutrality point, the bridge region becomes N-doped. In this bipolar gate configuration ($\Delta V_B > 0$), the QH edge states of the bridge region circulate in the opposite direction to those of the adjacent P-doped regions. Consequently, the QH edge states at a PN interface propagate in parallel, but are separated by a narrow tunneling barrier whose height and width are set by the $\nu=0$ QH state of the gapped BLG (Fig. 1C). We also note that the QH edge states with different carrier type are brought close at the PN junction that is away from the physical edges of the device. Without such careful engineering of QH pathways, one would expect tunneling 'hot' spots of QH edge states at the corners of a PN junction where branching occurs *(20, 21)*. Such an effect will lead to Aharonov–Bohm effect instead of quantum tunneling *(20, 21)*. By ensuring a constant tunnel barrier width and height along the PN junction, we have engineered a 1D interface that can be used to measure the spectroscopic signatures from tunneling between co-propagating PN QH edge states. This is in sharp contrast to previously reported work which lacked such intricate but necessary electrostatic design. These results showed full equilibration between QH edge states that led to an incoherent summation of quantum resistance between the P and N regions *(22, 23)*.

The different QH edge state configurations can be investigated by performing magneto-transport measurements across the device. Fig. 1D shows a two-terminal resistance as a function of $\Delta V_B$ for several different fixed magnetic fields. As the sign of $\Delta V_B$ changes from negative to positive, the channel configuration turns from PPP into PNP. In the PPP configuration, we find well-defined plateaus corresponding to $\nu=1$ and 2 of the bridge region. As $\Delta V_B$ increases, the channel resistance increases steeply. In particular, across the PPP-PNP boundary ($\Delta V_B=0$), the resistance increases by more than ten-fold, indicating tunneling transport across a barrier.

We investigate the transition regime further in Fig. 2 A by plotting the conductance (to emphasize features in the PPP regime) and resistance (to emphasize features in the PNP regime) as a function of $\Delta V_B$ and magnetic field $B$. In the unipolar regime ($\Delta V_B<0$), we observe a series of QH plateaus, at a relatively low field of 2 T, corresponding to filling factions $\nu=1, 2, 3, 4$, and 5. In the bipolar regime ($\Delta V_B>0$), however, there is no observable quantized conductance. Instead, the transport in the device shows two distinct features: (i) In the low field regime $B< B_c \sim 4.5$ T, the resistance increases slowly. This is followed by a rapid increase of resistance for $B> B_c$. (ii) Near the critical field, we observe resistance oscillations that are approximately independent of $\Delta V_B$. These two features are closely related to the onset of tunneling across QH edge states, as we discuss in detail below.

For $B< B_c$, the magnetic length $l_B = \sqrt{\frac{h}{eB}}$ is longer than the tunneling barrier width $d_{IS}$ set by the $\nu \neq 0$ incompressible strips (Fig. 2D). Since the decay length of the Landau level (LL) wave function across the IS is on the order of $l_B$, the QH edge states running parallel to the PN boundaries tend to be highly equilibrated *(13)*. In this case, the total resistance across the PNP junction should simply be the sum of the QH resistances of each doped region: $R_e = \frac{e^2}{h}(-\frac{2}{\nu} + \frac{1}{\nu'})$, where ν' and ν are the filling fraction of the N and P regions, respectively. Note that ν' is positive and ν is negative by definition. Similar transport was observed previously *(22, 23)*, where full/partial equilibration occurred across the PN junction. Fig. 2B shows the measured device resistance $R$ as a function of $B$ at a fixed gate voltage $\Delta V_B = 5$ V (corresponding to the white dashed line in Fig. 2B). In the low field regime ($B<B_c$), we observe that $R \approx R_e$ as expected. However, as B approach $B_c$, $R(B)$ starts to deviate from $R_e$. For $B>B_c$, $R$ increases exponentially as shown in Fig. 2E, and tunneling transport between well separated QH edges dominates.

Further quantitative analysis can be made by considering the tunneling process between QH edge states across ISs of the width $d_{IS}$. The current in each QH edge state is carried by a compressible electronic state with spatially varying chemical potential $eV_N(y)$, where $y$ is the position along the direction of PN junction and $N$ is the index for the QH edge states. We consider successive tunneling between neighboring QH edge states. Current conservation in a small length $dy$ of the $N$th QH edge state leads to:

$$\frac{4e^2}{h} dV_N(y) = -\{\gamma_N[V_{N+1}(y) - V_N(y)] - \gamma_{N-1}[V_N(y) - V_{N-1}(y)]\} dy \quad [1],$$

where $\gamma_N$ is the tunneling conductance per length between the QH edge states (see SM). This equation serves as our master equation to describe the observed data.

In the low magnetic field limit $l_B > d_{IS}$, and strong tunnel coupling between the CS regions effectively smears out the delineation between IS/CS within each doped region. This leaves the tunneling across the $v = 0$ region to dominate, and the overall tunneling barrier takes on a finite value $\gamma_0$. As a result, each N and P region behaves as one compressible state with filling fractions $v'$ ($v$). In this limit, Eq. (1) can be solved analytically (see SM) and leads to a simple expression: $R = \frac{h}{e^2}(-\frac{2}{v}\frac{1}{1-e^{-\alpha L}} + \frac{1}{v'}\frac{1+e^{-\alpha L}}{1-e^{-\alpha L}})$, where $\alpha = \frac{\gamma_0 h}{e^2}(-\frac{1}{v} + \frac{1}{v'})$ and $L$ is the length of the PN boundary. Note that this solution correctly reproduces the fully-equilibrated case at the large tunnel coupling limit where $e^{-\alpha L}$ vanishes. By fitting this model (see equation 11 in SM) to the data, we obtained $\gamma_0 = 500$ $\Omega^{-1}$m$^{-1}$, which is consistent with our experimental parameters.

As $B$ approaches $B_c$, the tunnel coupling between neighboring CSs becomes weaker and the transport across the device occurs via a cascade of tunneling across the series of ISs (Fig. 2E). This crossover between semiclassical to quantum transport regime occurs when $\ell_B = d_{IS}$. From the physical geometry of the PN junction and the contributions from multiple IS/CS, we estimate $B_c \approx 4.5$ T (see SM for detail), in accordance with our experimental observation. In the higher magnetic field regime, $B>B_c$, Eq (1) leads to an exponential dependence of R on B, which we use to characterize the profile of IS regions (see SM).

We now discuss oscillatory features we observe in the $R$ vs $B$ plot near the transition regime, $B \lesssim B_c$ (i.e., (ii) above). In this transition between semiclassical and full quantum transport regimes, the only appreciable tunneling across the PN boundary occurs through the $v = 0$ region between ISs of the doped regions. Thus, the modulation of the tunneling current can be ascribed to the alignment of Landau level (LL) on the P and N sides (Fig. 3A). In this regime, oscillations in R should correspond to on- and off- resonances between the LLs across the PN junction. In our device, we can achieve full experimental control of the energy alignment of Fermi level, P-LLs, and N-LLs by varying the magnetic field and $\Delta V_B$. The magnetic field causes the two sets of LLs to move in opposite directions in energy while $\Delta V_B$ moves them in the same direction. The LL filling fraction of each doped region is given by $v_{N,P} = h n_{N,P}/eB$, where $n_{N,P}$ are the electron and hole densities in the N- and P- regions, respectively. DOS is periodic in $1/B_F$, where $B_F = h(n_N + n_P)/e$ is determined only by the *PN* junction barrier height and not by any region-specific carrier density. We have deliberately designed the experiment so that the carrier densities in regions #4,5, and 6 are simultaneously controlled by $\Delta V_B$, so that the sum $n_N+n_P$ stays constant in the scan, thus keeping the oscillation periodicity unchanged (Fig 3A). Fig. 3B shows a plot of *R* as a function of 1/*B* for several different fixed gate voltages. We observe that the oscillation in *R* is clearly periodical in 1/*B*. From the periodicity observed in Fig. 3B, we estimated $n_P+n_N = 2 \times 10^{12}$ cm$^{-2}$, in a good agreement with the density of carriers calculated from $\Delta V_B$ and the capacitive coupling $C_B$. In Fig 3C, we display a map of *R* as a function of both $\Delta V_B$ and 1/*B*. Here, the oscillation in *R* appears as horizontal streaks but with a slight downward tilt as $\Delta V_B$ increases. The lowering of the peak with $\Delta V_B$ is due to decreasing v=0 tunneling barrier as the displacement induced BLG band gap $\Delta$ decreases. From the observed slope $\partial B/\partial V_B$ of the resonant tunneling peaks, we can estimate $\Delta$ using $\frac{\partial \Delta}{\partial V_B} = \hbar e(m_e^{-1} + m_p^{-1})\frac{\partial B}{\partial V_B}$, where $m_{e,p}$ are the electron and hole effective masses in the BLG, respectively. Fig. 3D shows experimentally obtained $\Delta$ as a function of displacement field $D = \varepsilon(V_T-V_B)/d_{hBN}$, where $d_{hBN} = 182$ nm is the thickness of the hBN layer, $\varepsilon$ is the dieletric constant of hBN, and $V_T(V_B)$ are local top (bottom) gate voltages. These values are in accordance with previously obtained values (*10*).

While keeping the energy alignment of P- LLs and N- LLs constant with respect to each other, sweeping $\Delta V_B$ along the resonant tunneling peaks adjusts the Fermi level alignment, which should also yield oscillations in resistance. To demonstrate this, we plot in the inset of Fig. 3C the first derivative of the measured resistance *dR/dB* in the scan range denoted by the dashed box. Along the resonance lines, we observe that *dR/dB* oscillates with a periodicity of $\Delta v_n \sim 4$, which is precisely the LL degeneracy in the low magnetic field limit. This agreement confirms that our spectroscopic transport technique can probe in detail the properties of LLs. Further development in the spatial engineering of QH edge state will allow us to explore emergent phenomena, such as the recently discovered quasiparticle interference (*24*, *25*), using the spectroscopic tunneling transport technique we demonstrate in this work.

**Acknowledgments:** We thank A. Yacoby and B. Halperin for helpful discussions. The major experimental work at Harvard University is supported by the U.S. Department of Energy (grant DE-SC0012260). K.W. is supported by Army Research Office (ARO) Multidisciplinary University Research Initiative (MURI) (grant W911NF-14-1-0247). P.K. acknowledges partial support from the Gordon and Betty Moore Foundation's EPiQS Initiative (grant GBMF4543). K.W. and T.T. acknowledge support from the Elemental Strategy Initiative conducted by the Ministry of Education, Culture, Sports, Science and Technology, Japan. T.T. acknowledges support from a Grant-in-Aid for Scientific Research (grant 262480621) and a grant on Innovative Areas "Nano Informatics" (grant 25106006) from the Japan Society for the Promotion of Science. This work was performed, in part, at the Center for Nanoscale Systems (CNS), a member of the National Nanotechnology Infrastructure Network, which is supported by the NSF under award no. ECS-0335765. CNS is part of Harvard University. A part of device fabrication was done in Albany NanoTech Institute supported by the Semiconductor Research Corporation's NRI Center for Institute for Nanoelectronics Discovery and Exploration (INDEX).

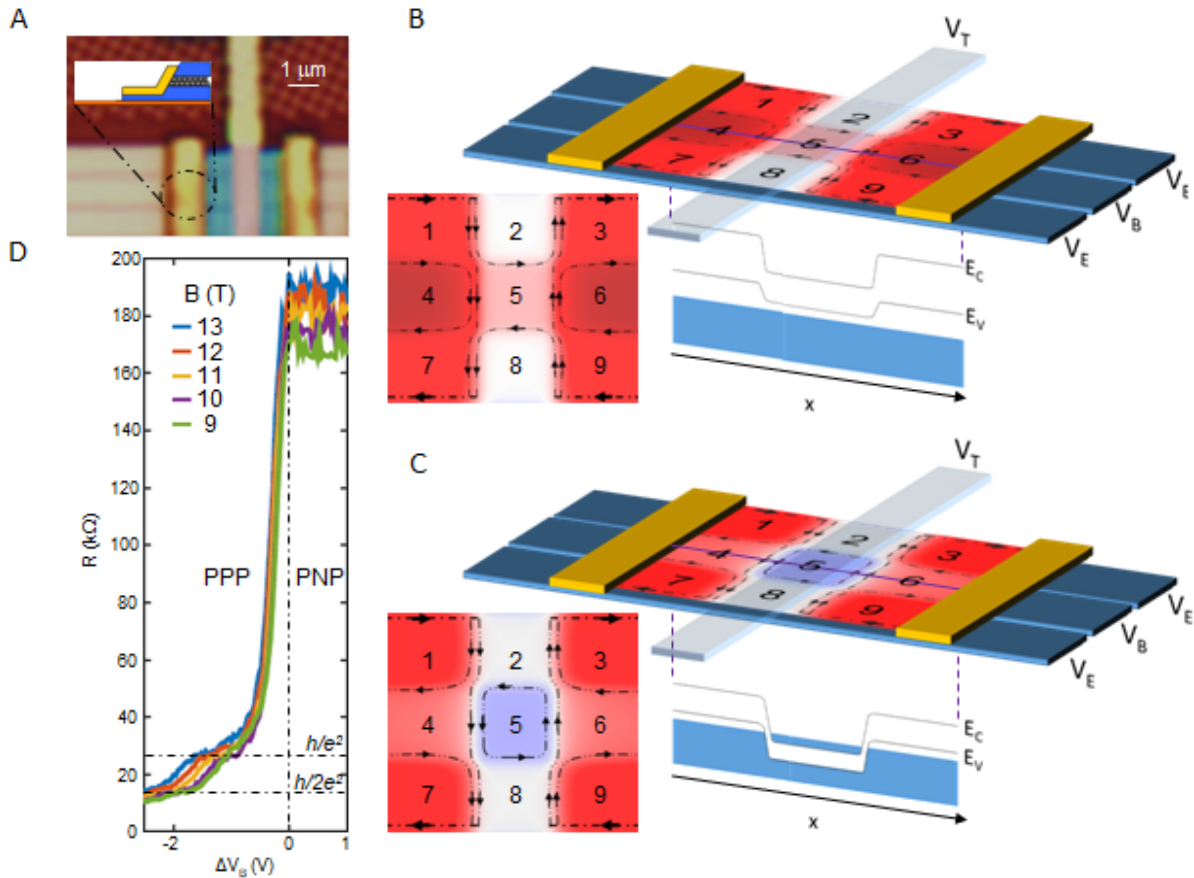

Figure 1. | **Gate-defned PNP Networks in Bilayer Graphene.** (A) Microscope image of the device, consisting of hBN-encapsulated bilayer graphene with three horizontal local backgates and one top gate. (inset) 1D edge contact are made with the bottom hBN half-way etched through. (B)(C) Top gate $V_T$ =8V and side backgates $V_E$=-10V are set to define insulating region #2 and #8. Only center gate $V_B$ are swept in the experiment, and we define $\Delta V_B$ = 0 at the charge neutrality point of center region #5 ($V_B$ = -10V). Simulated carrier density and edge state configuration at (B) $\Delta V_B$ =-5V and (C) $\Delta V_B$ =5V. The dashed lines depicts the spatial distribution of edge state current under magnetic field. When $\Delta V_B$<0 (B), current can flow through edge states whose number equals to filling factor of the center region of the device, and convectional quantum Hall transport behavior with quantized conductance plateaus is expected. However, when $\Delta V_B$>0 (C), the N-type center region (#5) is surrounded by a finite-width insulating region with a well-defined ν=0 Landau gap and displacement-field-induced bandgap. Magneto-transport across the sample is dominated by quantum tunneling and device enters non-equilibrium regime. (D) As a result, while quantized plateaus are clearly visible in the PPP regime ($\Delta V_B$<0), a steep change in measured magneto resistance occurs when transitioning to the PNP regime ($\Delta V_B$>0) .

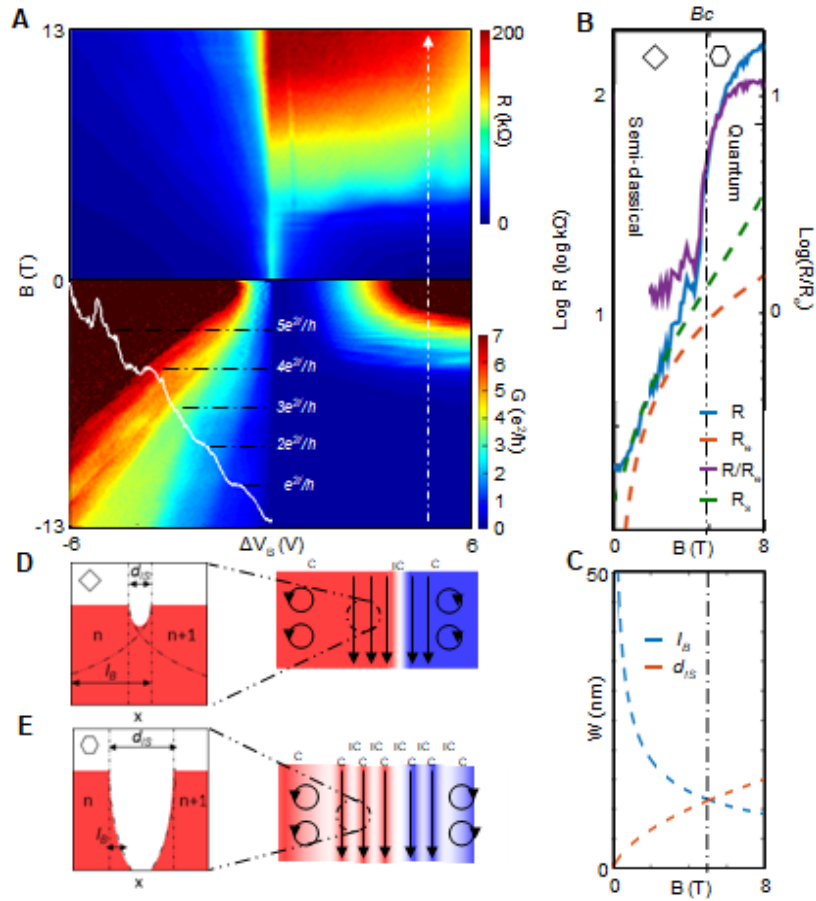

Figure 2. | **Quantum Hall Phase Transition in PN Configuration**. (A) Measured (top panel) resistance and (bottom panel) conductance across the device as a function of magnetic field B, While the fan diagram is visible in negative $\Delta V_B$, the device shows an unusually sharp magneto resistance rise around $B_C$ = 4.5 T at positive $\Delta V_B$. (B) Measured resistance ($R$), expected resistance with full equilibrium ($R_e$), fitting of the low field data using our analytical quantum tunneling model in semiclassical regime ($R_S$) and their ratio ($R/R_e$) as a function of magnet field sweep at $\Delta V_B$=5V (equal carrier density for P and N region, along the double dashed line in fig 2A). (C) Magnetic length $l_B$ and calculated incompressible strip width $d_{IS}$ as a function of field. (D) When $B<B_C$, the system is in the semiclassical regime. The magnetic length is longer than the incompressible strip width and therefore both the P and N side of the device behave like a single compressible state. Tunneling occurs only across the gate-defined v=0 tunnel barrier, which can be characterized by fitting the data using our analytical quantum tunneling model in this regime ($R_S$). (E) When $B>B_C$, the magnetic length becomes smaller than the widths of ISs and incompressible strips becomes well-defined due to reduced coupling between them. The cascade tunnelings are happening across all ISs, resulting in a steep double exponential increase in $R$ and $R/R_e$, and deviation from semiclassical description ($R_S$).

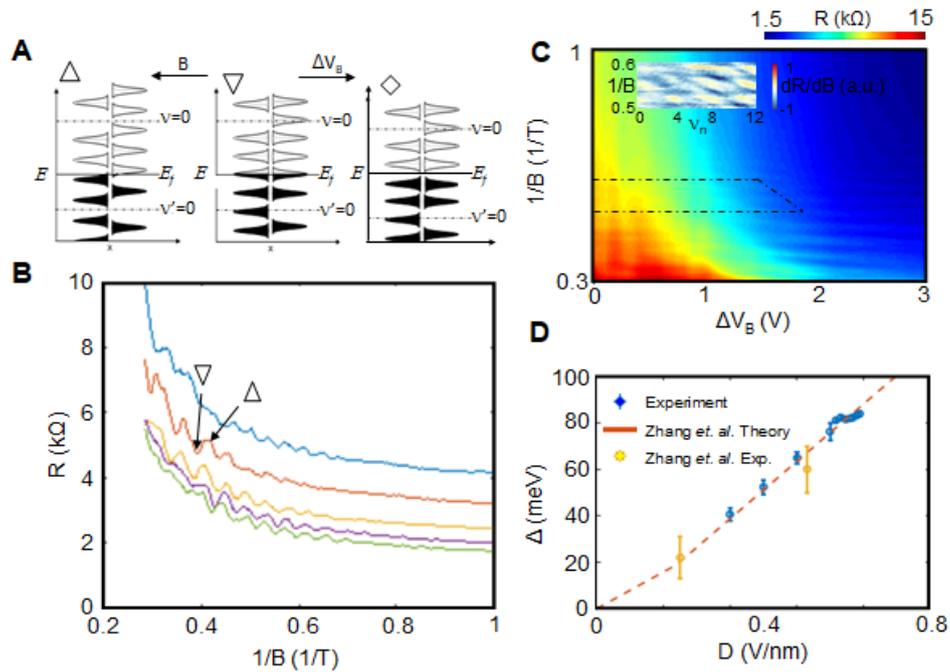

Figure 3. | **Resonant Landau Tunneling as a Spectroscopy Tool**. (A) In the semi-classical regime, the measured resistance depends on the alignment of Landau levels in the P and N side, as well as their alignment with the Fermi level. When sweeping the magnetic field, Landau level alignment changes from resonance (down-triangle) to off-resonance (up-triangle), periodically in inverse magnetic field. When sweeping $\Delta V_B$ while Landau levels are in resonance, the Fermi level can be temporarily unpinned and jump across the Landau gap (diamond), resulting into modulation of oscillation amplitude. (B) Measured resistance as a periodic function of 1/B at different gate configurations, and (C) 2D scan of the resonant tunneling features verifies that the Landau level alignments are only sensitive to the PN junction height instead of individual carrier densities in each respective region. (Inset) Differentiated data in the range of dashed region, a modulation in oscillation amplitude similar to that of Coulomb blockade has been observed with a periodicity of the electron filling factor $v_n$=4. (D) By tracing the peak positions as a function of field and gate voltage, we extract the spectroscopy bandgap of bilayer graphene as a function of displacement field D.